%% file: main.tex
\documentclass[runningheads]{llncs}
\usepackage{multirow}
\usepackage[font=small,labelfont=bf]{caption}
\usepackage{floatrow}

\usepackage[T1]{fontenc}
\usepackage{enumitem}
\usepackage{todonotes}
\usepackage[hidelinks]{hyperref}
\usepackage{graphicx}
\usepackage{comment}
\usepackage{algorithm}
\usepackage{algpseudocode}
\usepackage{comment}
\begin{document}
\title{Confidentiality-Preserving Verifiable Business Processes through Zero-Knowledge Proofs \\\vskip 1em \small{Preprint\thanks{This work has been accepted to be presented at the 29th International Conference on Enterprise Design, Operations, and Computing (EDOC 2025).}}}
\titlerunning{Preprint: ZKPs for Confidentiality-Preserving Verifiable Processes}
\author{Jannis Kiesel\inst{}\orcidID{0000-0002-7412-3746} \and
Jonathan Heiss\inst{}\orcidID{0000-0002-4239-8534}}

\institute{
Information Systems Engineering, Technische Universität Berlin, Berlin, Germany\\
\email{\{jaki,jh\}@ise.tu-berlin.de}}

\authorrunning{Preprint: J. Kiesel and J. Heiss}
\maketitle              %
\input{content/abstract}
\input{content/introduction}

\input{content/background}
\input{content/related_work}

\input{content/zk_applications}
\input{content/service_technology}
\input{content/evaluation}

\input{content/conclusion}

\bibliographystyle{splncs04}
\bibliography{main}
\end{document}

%% file: content/abstract.tex
\begin{abstract}

Ensuring the integrity of business processes without disclosing confidential business information is a major challenge in inter-organizational processes. 
This paper introduces a zero-knowledge proof (ZKP)-based approach for the verifiable execution of business processes while preserving confidentiality. 
We integrate ZK virtual machines (zkVMs) into business process management engines through a comprehensive system architecture and a prototypical implementation. 
Our approach supports chained verifiable computations through proof compositions. 
On the example of product carbon footprinting, we model sequential footprinting activities and demonstrate how organizations can prove and verify the integrity of verifiable processes without exposing sensitive information. 
We assess different ZKP proving variants within process models for their efficiency in proving and verifying, and discuss the practical integration of ZKPs throughout the Business Process Management (BPM) lifecycle. 
Our experiment-driven evaluation demonstrates the automation of process verification under given confidentiality constraints.

\keywords{Zero Knowledge Proof  \and Business Process Management \and Carbon Footprinting}
\end{abstract}

%% file: content/introduction.tex
\section{Introduction}

Inter-organizational business processes are fundamental to modern supply chains and collaborative ecosystems \cite{flynn_impact_2010,narayanan_antecedents_2011}. 
Yet they face a critical paradox: organizations require transparency to verify process integrity and build trust, while simultaneously needing to protect confidential business information from competitors and partners. 
This tension is particularly important in regulated domains such as carbon footprinting \cite{bhatia_product_2011}, healthcare data sharing \cite{fdhila_challenges_2022}, and financial compliance \cite{pisoni_blockchain-based_2023}, where regulatory frameworks increasingly mandate verifiable reporting while organizations cannot afford to expose sensitive operational data, proprietary methodologies, or competitive advantages. 
Traditional approaches to this challenge rely on either full disclosure to intermediaries, which creates confidentiality risks and limits scalability, or manual data validity checks. %
The result is a fundamental trust deficit that hinders effective inter-organizational collaboration \cite{panayides_impact_2009,muller_silver_2020} and forces organizations to either compromise on confidentiality or accept unverifiable claims from partners.

While the challenges of integrating confidential, yet verifiable, inter-orga\-nizational processes are relevant across various domains, this paper adopts the carbon footprinting issue as a representative use case.

\subsection{Running Example: Product Carbon Footprinting}
Product Carbon Footprints (PCF) quantify the Greenhouse Gas (GHG) emissions associated with a specific product throughout its lifecycle, from raw material extraction to disposal (cradle-to-grave) \cite{bhatia_product_2011}.  
By measuring and reporting these emissions, PCFs help manufacturers identify opportunities to reduce their environmental impact and enable consumers to make more carbon-conscious purchasing decisions.
Regulatory frameworks such as the EU Supply Chain Directive,\footnote{\href{https://eur-lex.europa.eu/eli/dir/2024/1760/oj/eng}{https://eur-lex.europa.eu/eli/dir/2024/1760/oj/eng}}
the Greenwashing Directive,\footnote{\href{https://eur-lex.europa.eu/eli/dir/2024/825/oj/eng}{https://eur-lex.europa.eu/eli/dir/2024/825/oj/eng}}
and the Corporate Sustainability Reporting Directive (CSRD)\footnote{\href{https://eur-lex.europa.eu/eli/dir/2022/2464/oj/eng}{https://eur-lex.europa.eu/eli/dir/2022/2464/oj/eng}} 
increasingly mandate reliable PCF reporting.

The GHG Protocol \cite{bhatia_product_2011} provides standards and tools to footprint GHG emissions.
It differentiates three organizational scope boundaries to attribute GHG-emitting activities: %
\begin{itemize}
    \item Scope 1 emissions are direct \textit{internal} GHG emissions from sources controlled by an organization, such as fuel combustion of manufacturing facilities. 
    \item Scope 2 emissions are indirect \textit{internal} emissions from purchased electricity, steam, etc., that an organization consumes. 
    \item Scope 3 emissions include all other indirect \textit{external} emissions in a company’s value chain, such as those from suppliers, product use, and waste disposal.
\end{itemize}

PCF standards mandate a business process activity-centric approach for calculating and tracking carbon emissions \cite{bhatia_product_2011}, which we refer to as \textit{footprinting}.
To calculate an accurate PCF, organizations must collect emissions data across a product's entire value chain, particularly for Scope 3 emissions, which account for the majority of emissions.
However, suppliers cannot provide full transparency in their calculation of PCFs, as disclosing sensitive business activity information may compromise confidential business process specifications~\cite{heiss_verifiable_2023}.
Due to the lack of transparency, suppliers must be assumed to greenwash their reported Scope 3 emissions.
This hinders trusted cooperation and requires organizations to verify the reported Scope 3 emissions.

\begin{figure}
    \centering
    \includegraphics[width=0.85\linewidth,trim={0cm, 0cm, 0cm, 0cm}]{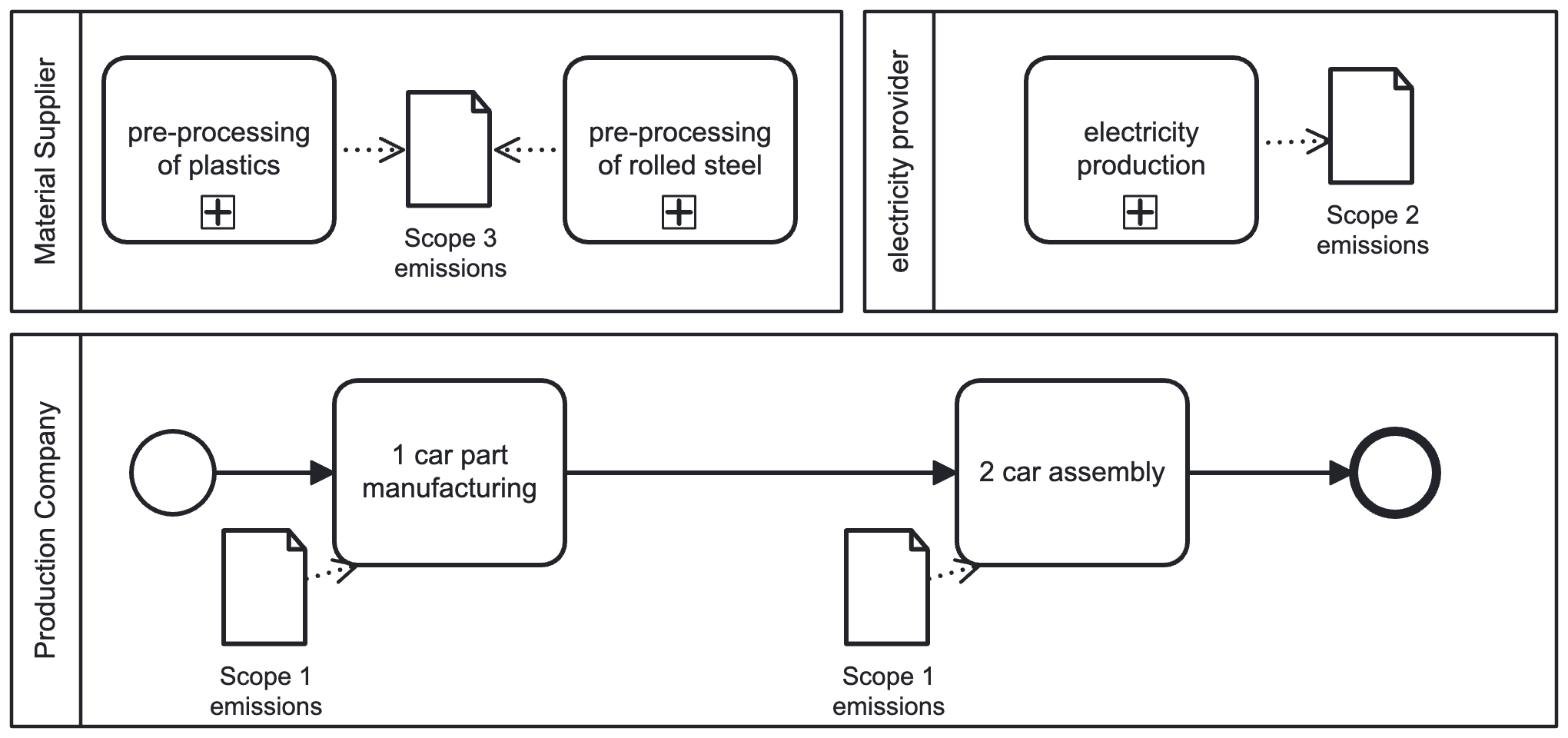}
    \caption{Car production process example for a cradle-to-gate process derived and reduced for clarity from the example process map of \cite{bhatia_product_2011}.} 
    \label{fig:PCF-example}
    
\end{figure}%

In the example shown in Fig. \ref{fig:PCF-example}, a material pre-processing supplier provides raw material inputs to a production company. 
To calculate a product's PCF, the production company relies on Scope 3 data from the supplier, Scope 2 data from their electricity provider, and Scope 1 emissions from the production process.

\subsubsection{Derived Requirements}\label{sec:requirements}

Key footprinting principles such as \textit{consistency} and \textit{transparency} must be met \cite{bhatia_product_2011}.
The underlying Business Process Management system is therefore required to facilitate these encompassing principles.
Therefore, the reporting enterprise's footprinting process must:

\begin{enumerate}
    \item \textit{Validate} that the Scope 3 emission data of all process inputs were computed based on compatible footprinting standards to ensure consistency,
    \item \textit{Prove} the PCF computation \textit{integrity} such that it can be verified to facilitate trust, and
    \item Ensure that sensitive process data, such as used electricity, materials, and process specifications, remain \textit{confidential} while making the footprinting \textit{transparent}.
\end{enumerate}

\subsection{Solution Approach}

Zero-knowledge proofs (ZKPs) verify the correctness of a computation without requiring access to the underlying inputs, utilizing cryptographic protocols. 
Recent advancements in ZKPs and related developer tools have been significant, mainly driven by blockchain applications, where ZKPs help address scalability and privacy issues. 
Beyond blockchains, ZKPs hold great promise for trustworthy and confidentiality-preserving inter-organizational data sharing~\cite{heiss_verifiable_2023}.
Despite their potential, ZKPs' cryptographic complexity and enterprises' unfamiliarity with the technology pose challenges for their adoption within processes.

This paper addresses the problem of trust and confidentiality in data outputs of business processes with the example of carbon footprinting and contributes in the following ways:
\begin{itemize}
    \item First, we outline confidentiality-preserving verifiable processes.
    \item Second, we present a service architecture to integrate ZKP technologies within business process engines.
    \item Finally, we evaluate the performance and feasibility of our implementation.
\end{itemize}

Our contributions unfold as follows: We provide background information in Section \ref{sec:background} and related work in Section \ref{sec:related-work}. Next, we present ZKPs for confidentiality-preserving verifiable processes in Section \ref{sec:carbon-footprinting}.
Afterward, we present our service architecture implementation in Section \ref{sec:service-architecture}.
Finally, Section \ref{sec:evaluation} evaluates our proposed solution, and Section \ref{sec:conclusion} concludes.

%% file: content/background.tex
\section{Background}\label{sec:background}

Below, we summarize the foundation upon which our work is built.

\textbf{Business Process Management:}
Business processes consist of coordinated activities executed within an organizational and technical environment by a single organization to achieve a specific business objective. 
They often involve interactions with processes from external organizations. 
Using the Business Process Management (BPM) lifecycle, encompassing management concepts, methods, and techniques, these processes can be efficiently structured, standardized, repeated, and automated \cite{WeskeMathias2019BPMC}. 
In orchestrator-based Workflow Management Engines (WMEs), atomic process activities are executed through distributed, horizontally scalable service tasks.
The WME manages the propagation of activity data, while event data, generated by application systems or business process execution platforms, is recorded in event logs. 
Camunda\footnote{\href{https://www.camunda.com}{camunda.com}} is a widely used and extensible open-source WME that features a built-in orchestrator-worker architecture.

\textbf{Zero-Knowledge Proofs:}
ZKPs describe a class of protocols that enable one party (\textit{prover}) to convince another party (\textit{verifier}) of the correctness of a statement without disclosing private arguments. This adds the notion of ``zero knowledge.'' Non-interactive ZKPs furthermore allow this with a single message. While ZKPs were invented in the 80s \cite{goldwasser_knowledge_1985}, they have received immense attention in recent years for their suitability to mitigate the confidentiality and scalability limitations of blockchain technology \cite{mendling_blockchains_2018,muller_silver_2020,weber_untrusted_2016}.
For that, zkSNARKs, as described in \cite{gennaro_quadratic_2013}, are particularly suitable, which allow the verification of arbitrary large computations at constant costs (succinctness). 
STARKs, as introduced in \cite{ben-sasson_scalable_2018}, 
address some limitations of zkSNARKs, removing trust assumptions from the initial setup of zkSNARKs and adding post-quantum security~\cite{eberhardt_offchaining_2018}. %

\textbf{Zero-knowledge Virtual Machines}:
STARKs serve as the foundation for \textit{zero-knowledge virtual machines} (zkVMs), which provide developers with tools to build ZK-enabled applications without needing to understand the underlying cryptographic complexities.
In contrast to application-specific circuits like e.g. ZoKrates~\cite{eberhardt_zokrates_2018}, zkVMs encode a virtual machine's instruction set as a circuit \cite{dokchitser_zero_2023}. 
ZkVMs translate programs written in common programming languages like Rust into a suitable binary format and execute it within virtual machines (VM) with zk-instructions natively supported in the VM’s instruction set. 
Apart from the outcome of the original program, zkVMs deliver a ZKP as proof of correct execution. 
They provide a convenient tool for specifying code using higher-level programming languages and abstract many security and performance issues.

\textbf{Risc0:}
In this work, we leverage Risc Zero (Risc0)\footnote{\href{https://risczero.com}{risczero.com}} as an established zkVM available at the time of writing which utilizes zkSTARKs. Risc0 is built upon the RiscV instruction set,\footnote{\href{https://riscv.org}{riscv.org}} which is an open-source standard for VMs and, hence, promises interoperability with many programming languages. 
The procedure for Risc0 can be described in the following steps \cite{bruestle_risc_nodate}:

\begin{enumerate}
    \item \textit{Compilation}: The prover specifies a program in Rust, called the \textit{guest program}, which is compiled into an executable format for the RISC-V instruction set. Each method in a \textit{guest program} is compiled into an Executable and Linking Format (ELF) binary before execution begins.
    \item \textit{Execution}: The executor runs the ELF binary and records the session, which is the execution trace, i.e., a complete record of the computation containing complete snapshots of the state of the zkVM at given moments in time. 
    \item \textit{Proving}: The prover checks and proves the validity of the session, outputting a \textit{receipt}. The \textit{receipt} serves as a succinct proof of validity for executing the application. Receipts can be passed to third parties and verified to attest to the cryptographic validity of the application’s output.
    \item \textit{Verification}: The verification algorithm receives an \textit{ImageId} as a parameter; the \textit{ImageId} serves as a cryptographic identifier for the expected ELF binary.
\end{enumerate}

%% file: content/related_work.tex
\section{Related Work}\label{sec:related-work}

We consider the related work on trust-enhancing approaches for choreographies and ZKP approaches within BPM.

Blockchain platforms have primarily addressed the issue of trust and verifiable computing between organizations.
Enacting collaborative business processes on permissioned blockchains was first proposed by Weber et al. \cite{weber_untrusted_2016} by translating BPMN process models to smart contracts and was extended in 
\cite{lopez-pintado_caterpillar_2017,garcia-banuelos_optimized_2017}. 
Sturm et al. \cite{sturm_lean_2019} proposed a system that enables on-chain execution of business processes using generic smart contracts populated with specific process models. 
By storing the process specification directly on-chain, their approach improves transparency and trustworthiness compared to earlier solutions, such as Weber et al. \cite{weber_untrusted_2016}. 
However, this increased transparency comes at the cost of confidentiality, as all participants must agree on and expose a globally shared process specification, making the approach unsuitable for scenarios involving sensitive data or proprietary logic.

To preserve the confidentiality of private process data in a public blockchain environment
Marangone et al. \cite{marangone_fine-grained_2022} adopt an Attribute-Based Encryption (ABE) approach and extend it to a Multi-Authority ABE approach \cite{marangone_martsia_2024}. 
While their method effectively protects sensitive data contents, it still requires disclosure of process specifications and metadata.

For concealing process states, Toldi and Kocsis \cite{toldi_blockchain-based_2023} employ ZKPs and the ZoKrates toolkit \cite{eberhardt_zokrates_2018} to implement “zero-knowledge workflows,” in which encrypted process states derived from BPMN models are stored on-chain to enable confidential collaboration across organizations.
While their approach successfully conceals private process data, it suffers from high on-chain computational costs and limited scalability, rendering it impractical for real-world deployment. 
Moreover, it exposes process specifications and metadata to other parties. 
Building on Toldi and Kocsis, Petto et al.~\cite{petto_interpreted_2024} significantly enhance efficiency and eliminate the need for process-specific smart contracts by embedding the process model as a private input within the ZKP, thereby ensuring the confidentiality of the process specification. 
ZKPs also find adoption in service-oriented approaches for trustworthy and privacy-preserving data exchanges across organizations in the context of collaborative business intelligence~\cite{quattrocchi_trustworthy_2024}, IoT data pre-processing~\cite{heiss_trustworthy_2021}, cloud-native scalability~\cite{servicifying_snarks}, or chained trusted compute units~\cite{castillo2025trusted}.

Concealing sensitive data and process internals, even utilizing encryption, remains a challenge on-chain as other organizations can observe which accounts interact and how often with whom \cite{weber_untrusted_2016}. 
Solution approaches to this carry their own issues, such as new account addresses for each process instance hindering the possibility of creating reputable accounts \cite{weber_untrusted_2016}.
To the best of our knowledge, we are the first to introduce a blockchain-less architecture that supports scalable, verifiable, and confidential process execution without exposing any process data or metadata to other parties.

%% file: content/zk_applications.tex
\section{Confidentiality Preserving Verifiable Processes}\label{sec:carbon-footprinting}
In line with the running example and derived requirements of Section~\ref{sec:requirements}, we introduce a confidentiality-preserving verifiable process model starting with a baseline model, and then extend it through validity and integrity. 

\begin{figure}[b]
    \centering
    \includegraphics[width=1\linewidth,trim={0cm, 2cm, 0cm, 2cm}]{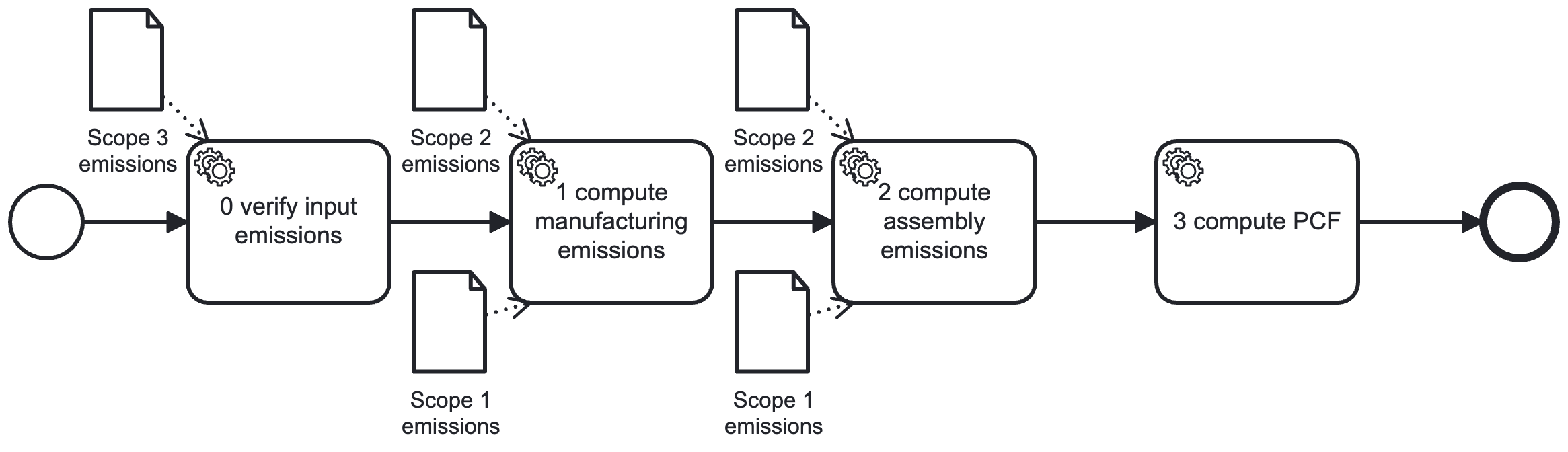}
    \caption{Footprinting process for the car production process of Figure \ref{fig:PCF-example}. It resembles the composite verifiable process strategy outlined in \ref{sec:a2}.} 
    \label{fig:footprinting_process}
\end{figure}

\subsection{Baseline Process}
Consider a simplified footprinting process for the PCF of a car as shown in Figure \ref{fig:footprinting_process}.
The footprinting process consists of four activities. First, in 0, the Scope 3 emissions from the material supplier are validated. 
Then, in 1 and 2, the manufacturing emissions and assembly emissions are calculated using the internally assessed Scope 1 direct emissions and Scope 2 emissions provided by the electricity provider.
Finally, in 3, the Scope 1, 2, and 3 emissions of the process are aggregated to the total PCF.

The validation in 0 faces the following problem: To validate the Scope 3 emissions of the supplier, the production company must have access to the supplier's sensitive business activity data to validate that the footprinting was carried out according to a compatible standard and verify that the Scope 3 emissions were computed correctly.
In this baseline process, verification and validation must be conducted manually by an emission footprinting expert, which increases the process's cost and decreases its performance as it cannot be fully automated.
However, the supplier cannot provide confidential business information necessary for the PCF validation to the production company.
In the absence of process transparency, the production company must trust in the reported Scope 3 emissions validity and computational integrity.

\subsection{Adding Validity}\label{sec:adding_validity}
\begin{figure}[t]
    \centering
    \includegraphics[width=0.7\linewidth]{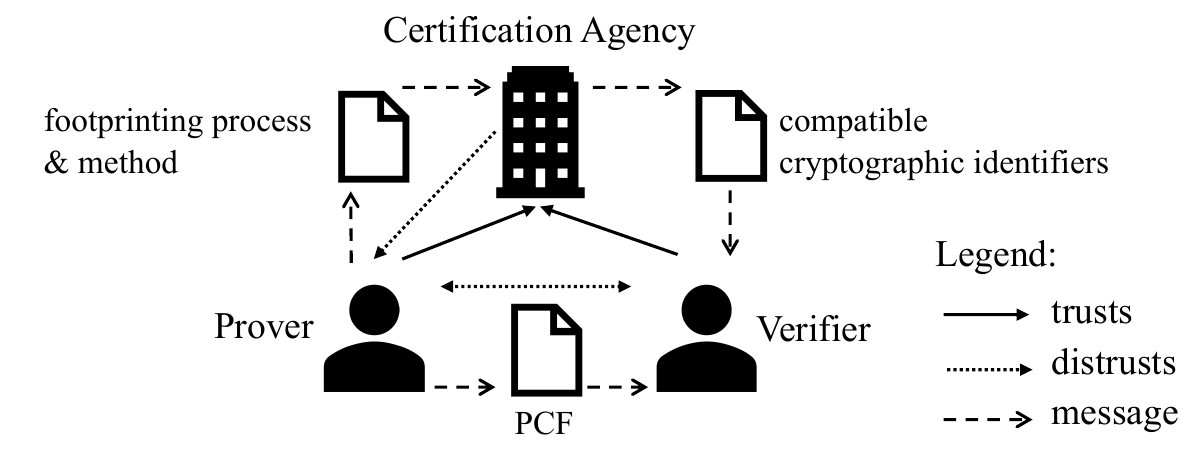}
    \caption{Trust relationships and artifact exchange.}
    \label{fig:trust-assumptions}
\end{figure}

To resolve the conflict between transparency and confidentiality, we introduce a footprinting certification agency which acts as a trusted third party among all process participants.\footnote{Utilizing a certification agency aligns with industry standards for PCF processes as outlined in the Catena-X Product Carbon Footprint Rulebook, \href{https://www.carbon-transparency.org/resources/catena-x-product-carbon-footprint-rulebook}{carbon-transparency.org/resources/catena-x-product-carbon-footprint-rulebook}.} %
In Figure \ref{fig:trust-assumptions}, the supplying company is the \textit{Prover} and the production company is the \textit{Verifier}.

Before the process is enacted, the proving party discloses its footprinting process and calculation method to the certification agency, which audits the process and method. 
It attests conformance of the process and method with a PCF standard through a signed footprinting certificate, which is returned to the proving party.
Whenever the footprinting process changes, the certification agency performs a new audit and certifies the validity of the PCF standard for the changed process.
On process enactment, the proving party sends their reported Scope 3 emission data, along with the footprinting certificate, to the verifying party.
The verifying party trusts the certification agency and therefore only needs to validate the compatibility of the Scope 3 emission certificate with its footprinting process.

Sharing the footprinting process with the certification agency facilitates the transparency principles of PCF standards. 
At the same time, the private process activities of the prover and their data remain confidential from other parties as they are not shared.

\subsection{Adding Integrity}

The supplier can alter the execution of the footprinting method to greenwash their PCF even if the footprinting process was audited and certified by a trusted third party.
Because the supplier does not disclose its footprinting process and method to the production company, the production company must trust in the supplier's footprinting integrity.

To remove this trust assumption, we propose to utilize ZKPs for all footprinting methods that require verification by other parties.
The footprinting process activities of the supplier are executed from within zkVM guest programs.
Shown in Figure \ref{fig:trust-assumptions}, the footprinting process and footprinting method, in the format of a zkVM guest program, are shared with the certification agency.
The certification agency generates the proving and verification keys for the guest program of the footprinting process. 
It provides the proving key to the prover and the verification key to the verifier.%

Upon process enactment, the prover executes the footprinting guest program, proves its computational integrity using the proving key, and sends the proof, which includes the Scope 3 emissions, to the verifier.
The ZKP guarantees the integrity of the computation of the Scope 3 emission data.
The verifier, in activity 0 of Figure \ref{fig:footprinting_process}, verifies the ZKP using the verification key.
Because the verifier trusts in the footprinting validation of the certification agency, verifying the ZKP using the certified verification key also validates that the footprinting process is consistent with the PCF standard.

In turn, the production company must also prove the PCF of the assembled car to the customer, and cannot disclose confidential business information. 
Therefore, for the remaining footprinting process, the production company assumes the role of the prover and executes all its footprinting activities within zkVM guest programs, which are also certified by the certification agency. 
In 1 and 2, the prover verifies the Scope 2 emissions from the electricity provider and computes and proves the Scope 1 emissions of the production activities.
Finally, in 3, the prover aggregates all footprinted emissions from previous footprinting activities and creates a ZKP that contains the total PCF of the car.

%% file: content/service_technology.tex
\section{Integrating zkVMs into Workflow Management Engines}\label{sec:service-architecture}
We propose extending WMEs with zkVM-enhanced tasks to perform verifiable computations of activities.

\subsection{System Architecture}
\begin{figure}[t]
    \centering
    \includegraphics[width=0.85\textwidth,  trim={2cm, 4.5cm, 8.5cm, 0cm}, angle=0]{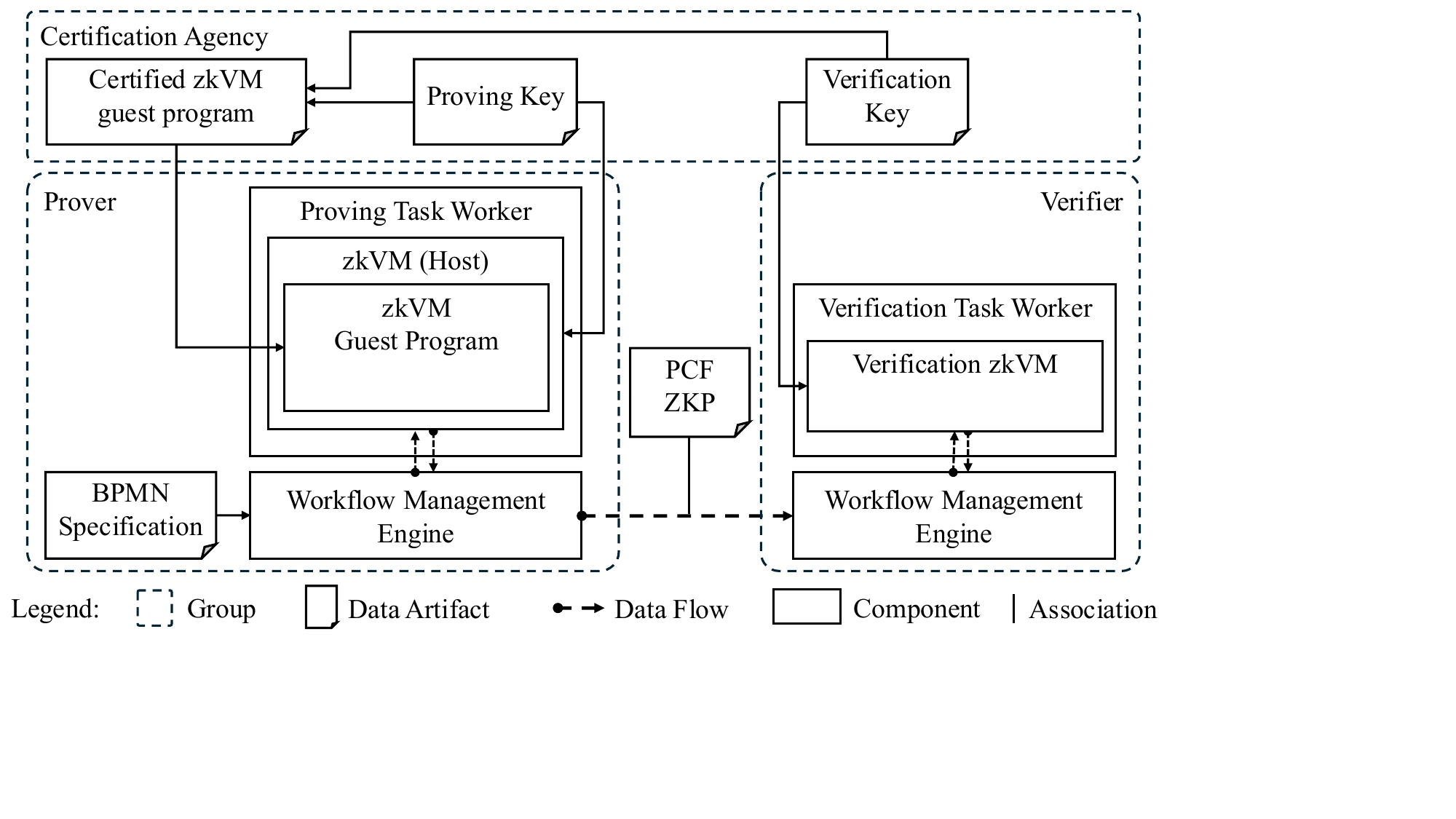}
    \caption{Architecture for extending WMEs with zkVM task workers.}
    \label{fig:service-architecture}
\end{figure}

Our architecture comprises the prover and verifier with their respective WMEs and dedicated modular proving and verification task workers as shown in Figure \ref{fig:service-architecture}.\footnote{Our architecture implementation artifact within the Camunda Platform is open-source available here: \href{https://github.com/curiousjaki/verifiable-processes-demo}{github.com/curiousjaki/verifiable-processes-demo}}
A proving task worker contains the proving zkVM host and the zkVM guest program. 
In contrast, the verification task worker consists only of a verification zkVM host with the verification keys of the trusted certification agency.
The WME orchestrates the verifiable process, handles errors and logs, and propagates the resulting ZKPs between the proving and verification activities as process variables. 
Using the service task approach, our proposal leverages horizontal scaling orchestrated by the WME or an external load balancer.
Generally, utilizing zkVMs within business processes can be distinguished into two phases: setup and operation.

\textbf{Setup:}
In the setup phase, the certification agency audits and certifies the \textit{provers'} guest program and generates the proving and verification keys.
The verification key is shared with the verifier.

\textbf{Operation:}
Upon executing a verifiable process, the \textit{provers'} WME calls the zkVM host to run the certified guest program in a guest zkVM. The zkVM host provides the private and public data inputs to the guest program.
In the context of our running example, the footprinting guest program receives the private business data, such as resource amounts and public emission factor data. The guest program then returns the PCF as part of the public data output of the guest program.
The zkVM host then creates a computational integrity proof of the guest program using the proving key, resulting in a ZKP that includes the public data output of the guest program.
The ZKP size and proving runtime are determined by the complexity of the guest program and the size of the data input, and are constrained by the WME and its variable data store size.
Upon successful proof generation, the ZKP is returned to the WME and then propagated to the verifying party.
The \textit{verifiers'} WME then sends the ZKP to a verification task worker. The verification zkVM determines the correct verification key, verifies the presented ZKP, and extracts the public data output.

\subsection{BPM Lifecycle}\label{sec:lifecycle}

Integrating ZKPs within existing BPM practices allows the utilization of the managerial aspects of the BPM lifecycle for ZKP management.
In the following, we map ZKP procedures to BPM lifecycle phases outlined around the lifecycle of Weske \cite{weske_business_2012} with its Design \& Analysis, Configuration, Enactment, and Evaluation phases.
The proving and verifying participants align their verifiable process \textit{design} and \textit{configuration} lifecycle phases with the certification agency as described in Section \ref{sec:adding_validity}.
In the design and analysis phase, the proving participant identifies external process data inputs that need to be verified before internal processing occurs and discloses their proving \textit{guest programs} with the certification agency.
The validity of the \textit{guest program} is then manually assessed and the proving and verification keys are generated.
In the configuration phase, task-specific zkVM proving and verification \textit{guest programs} are configured as part of the business process.
Each participant implements their own proving and verification system architecture utilizing compatible zkVM systems.
The enactment phase is extended such that the proving participant executes the defined business process and sends the \textit{guest programs} resulting ZKP to the verifying participant.
The verifying participant handles verification errors.
In the evaluation phase, the performance of the proving and verification tasks is evaluated based on their effects on relevant quality dimensions and their relevance to the process control flow.
Moreover, new private data inputs and outputs are identified for the following iteration of the BPM lifecycle.

%% file: content/evaluation.tex
\section{Evaluation}\label{sec:evaluation}

We quantify the computational overhead of integrating zkVMs in processes through an experiment, evaluate the confidentiality and transparency requirements of Section \ref{sec:requirements} qualitatively, and discuss threats and validity limitations.

\begin{figure}[h]
    \centering
    \includegraphics[width=0.7\linewidth, trim={1cm, 5cm, 14cm, 0cm}]{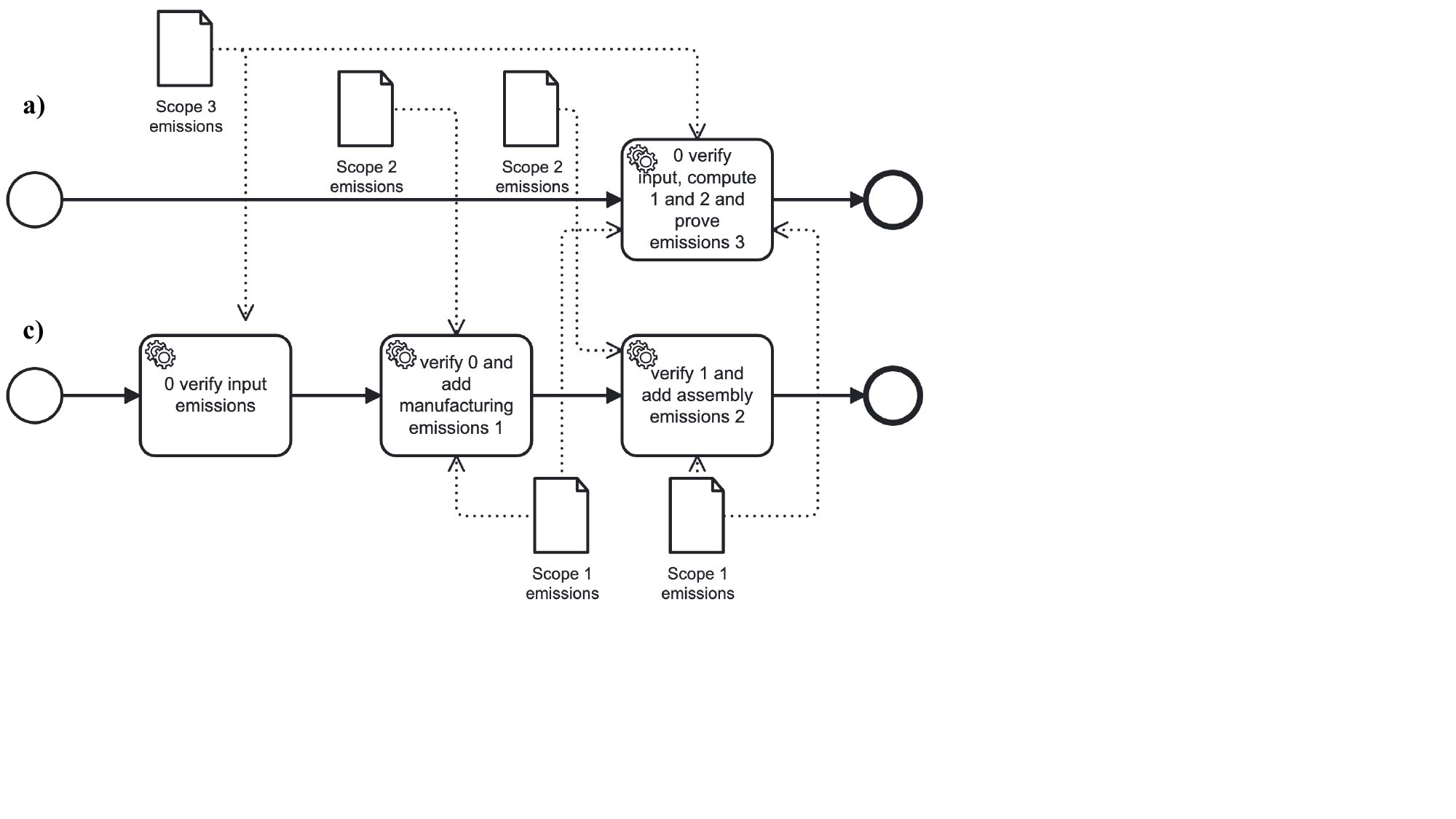}
    \caption{Verifiable process modeling alternatives \ref{sec:a1} single-step and \ref{sec:a3} chained for the PCF example process of Figure \ref{fig:PCF-example}.}
    \label{fig:PCF-example-2}
\end{figure}

\subsection{Modeling Alternatives for Verifiable Processes}\label{sec:modeling_alternatives}

We investigate three alternatives to the carbon footprinting process approaches visualized in Figure \ref{fig:footprinting_process} and Figure \ref{fig:PCF-example-2}:
\begin{enumerate}[label=\textbf{\alph*)}]
    \item \textbf{Single-step proving:}\label{sec:a1} All verifiable processing steps are carried out within one verifiable activity.
    In our running example, the single activity verifies Scope 3 and Scope 2 emissions, computes the Scope 1 emissions of all business activities, and proves the computational integrity of its footprinting in a single ZKP with its public PCF output.
    \item \textbf{Composed proving:}\label{sec:a2} A verifiable activity proves a single business activity, with a final activity composing all prior proofs.
    This approach is the example footprinting process outlined in Figure \ref{fig:footprinting_process}.
    This composing approach introduces the notion of process-centricity as the different footprinting activities prove individual business activities.
    The process notion enables the assessment of carbon footprints on a per-activity basis, allowing for the analysis of ZKP performance through a process mining lens.
    \item \textbf{Chained proving:}\label{sec:a3} Altering the composed approach, individual verifiable activities can be chained, where each verifiable activity verifies the prior proof and aggregates the prior's public data output with its computation result.
    In our example, the verification activity 0 generates a proof containing the Scope 3 emissions. The footprinting activity 1 verifies the proof of activity 0, and computes its business activity footprint and aggregates it with the output of the proof of 0.
    Then, 2 verifies the proof of 1, computes its business activity, and aggregates it with the public data output of 1 and so on, omitting the need for a final aggregation step.
    This approach is suitable for processes that may not entirely complete or exit early due to errors, as a cumulative PCF of the enacted activities is continuously present.
\end{enumerate}

\subsection{Verifiable Carbon Footprinting Guest Program}

In our chained footprinting example of Figure \ref{fig:PCF-example-2}, each footprinting activity computes an activity's Scope 1 emissions in a dedicated zkVM \textit{guest program} as illustrated in Algorithm \ref{alg:example}.
The proving method processes private data such as the emission factors and resource amounts.
As the \textit{guest program} is executed within a zkVM, its outputs are encoded within a ZKP generated by the zkVM host.
The ZKP includes the public PCF computation outputs required by the footprinting standard and ensures the integrity of the program's computation.

\input{content/computation_algorithm}

With our \textit{chained} approach, we designate the outputs of prior computations as inputs of successive proving and verification activities.
The \textit{chaining} is facilitated within a respective \textit{guest program} through proof composition, where the ZKP of a previous activity is provided to the zkVM guest as private data input.
The \textit{guest program} thereupon verifies the ZKP within the zkVM host and encodes the verification success within its own ZKP.
The resulting outputs include the cumulative emissions of the entire process instance and the specific emissions of the current activity. 
Thus, the last footprinting activity output becomes the PCF data output of the footprinting process, which is then used as input for successive footprinting processes of verifying parties, as Scope 3 emission data.
The \textit{composed} approach utilizes the same composition approach but in a dedicated final composition \textit{guest program}.
The proof composition of the \textit{chained} and \textit{composed} approach enables the efficient verification of the entire footprinting process.
A verifying party merely needs to verify the ZKP of the last footprinting activity to ensure the validity and computational integrity of the prover's footprinting process.

\subsection{Quantitative Evaluation}
We have implemented all the approaches presented in Section \ref{sec:modeling_alternatives} to evaluate their practical feasibility on the example of Risc0 in the popular Camunda WME.\footnote{\href{https://camunda.com}{camunda.com}}
We chose Risc0 for its mature state, open-source availability, and rich development support, and opted for Camunda for its wide adoption, extensibility, and inherent orchestrator-worker architecture.
For Risc0 zkVMs, the proving and verification keys are the same, referred to as \textit{ImageId}.

We have generated processes of start and end events and activities connected with message flows, omitting gateways and other BPMN elements.
In our experiment, we created processes with one to thirty activities to measure how process length impacts footprinting proving performance.
Our footprinting \textit{guest program} multiplies a resource input with an emission factor, generating a carbon footprint, which is then aggregated with a previously calculated PCF as outlined in Algorithm \ref{alg:example}.
All experiments were conducted on an 8-core Apple M3 processor with 16GB of LPDDR5 memory, utilizing a Risc0 zkVM version 1.2.5.

\newfloatcommand{capbtabbox}{table}[][\FBwidth]
\begin{figure}[t]
\begin{floatrow}
\capbtabbox{%
  \input{media/table}
  \vspace{-2em}
}{%
  \caption{Process proving overhead in sec and proof size in MB.}%
  \label{tab:results}
}

\ffigbox{%
  \includegraphics[width=0.4\textwidth, trim={0.5cm, 0.5cm, 0.5cm, 0.5cm}]{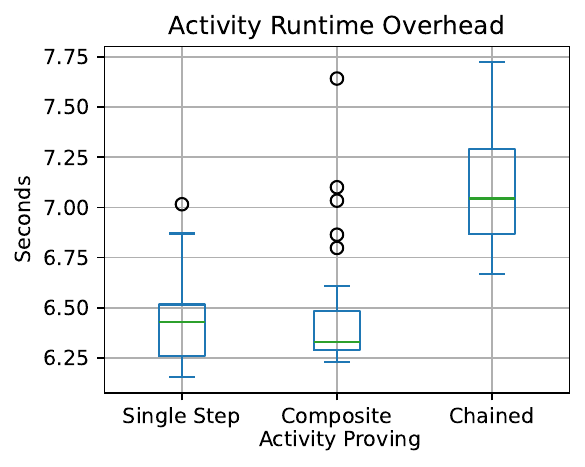}%
}{%
  \caption{Proving runtime overhead of an\\ activity within a process by approach.}%
  
  \label{fig:results}
}
\end{floatrow}
\end{figure}

For the \textit{single step} approach, only one proving activity is carried out with varying resource input variables proportional to the number of activities within the process.
Our results of Table \ref{tab:results} show that we could not observe the effect of the number of variables passed into the \textit{guest program} on the proof size and proving runtime overhead, suggesting near-constant costs for each process instance.

In the \textit{composite} approach, the proving activities are followed by one final composing activity, which is not required in the \textit{chained} approach as it composes the previous receipt in each proving activity.
Table \ref{tab:results} shows that the generated receipts and proving runtime overheads of the \textit{composite} and \textit{chained} approach increase linearly with the number of activities verified within the process.
The \textit{composite} approach yields a larger proof size due to our recursive implementation of the \textit{composite} proof aggregation method.
Other composition methods yield proof sizes comparable with the \textit{chained} approach but require specific composition \textit{guest programs} per number of activities of a given process.

As shown in Figure \ref{fig:results}, the proving overhead of the \textit{chained} approach per activity is higher than the composite and single-step approaches, caused by the proof chaining in each activity. 
However, it is more efficient on a process level than the \textit{composite} approach as shown in Table \ref{tab:results}.
The overhead of the final composition step in the \textit{composed} approach depends on the number of previous receipts to verify and compose, which decreases the overall process instance performance.
It is less efficient than the \textit{chained} approach at three activities and doubles the overhead of the \textit{chained} approach at 30 activities.

For a carbon footprinting process involving 30 business activities without proof generation, the execution time is 0.06 seconds for the single-step approach, 1.90 seconds for the composed approach, and 2.19 seconds for the chained approach. 
With proof generation, the process instance runtimes are much larger. In the composed approach for example, the total process runtime was 465.94 seconds and the proof generation accounted for 99.6\% of the total runtime, highlighting the proving overhead of the zkVM.
In contrast, the verification duration of receipts increases linearly with the proof size, rather than with the approach, ranging from 0.04 seconds for a small proof of a process involving one activity to 0.73 seconds for verifying a composite proof of 30 activities.

Our results show that the different approaches entail significant performance and feasibility constraints.
Messaging standards such as gRPC and WMEs such as Camunda have message and variable size limitations of 4MB.\footnote{\href{https://docs.camunda.io/docs/components/concepts/variables/\#variable-size-limitation}{docs.camunda.io/docs/components/concepts/variables/\#variable-size-limitation}}
This constraint prevents the propagation of proofs by WMEs for processes with a larger number of activities.
In our implementation with Risc0, we reached this limit in our \textit{composite} and \textit{chained} approaches at 8 and 16 activities, respectively.

\subsection{Qualitative Evaluation}

As with blockchain approaches, where enterprises must agree on a blockchain platform, our ZKP approach requires enterprises to agree on the zkVM vendor and version.
In contrast to blockchain, computation remains within each participant's organizational boundaries, allowing participants with different confidentiality concerns to collaborate without an all-in commitment.

Our different PCF approaches show tradeoffs in their performance and applicability to real-world scenarios.
The \textit{single step} approach benefits from near-constant proving runtime overheads and proof sizes and, therefore, the highest integrability in standard WMEs.
However, it only creates a verifiable computation at the end of a process, thereby prohibiting process management tools from managing the footprinting.
The \textit{composite} approach performs worse than the \textit{chained} approach for proof size and instance duration.
Both enable process-centric footprinting.
The chained approach additionally ensures that the footprinting is \textit{consistent} throughout at runtime, leveraging the integrity guarantees of zkVMs. %
Despite the proving and composition overheads, we argue that the \textit{single-step} and \textit{chained} approaches are viable for process-centric verifiable PCF between organizations due to their guarantees of confidentiality and efficient verifiability.
Alternative PCF approaches require full processing transparency between participants to ensure validity, or involve time-consuming manual spot-checking, and do not allow for complete verification of PCF computation integrity.

While our evaluation focuses on carbon footprinting as a representative use case, the proposed approach is equally applicable to other regulated domains such as health data sharing and financial compliance, thanks to our system's adaptability to varying data models and validity requirements.

This work entails several limitations.
First, our evaluation process is relatively simple, which may limit the generalizability of the results to more complex process models.
Furthermore, we do not evaluate the usability or organizational adaptability of our proposed implementation, both of which are crucial for practical adoption.
Specifically, our static zkVM setup does not account for dynamic process changes that may occur in practice.

\subsection{Degree of Confidentiality}
Similar to Petto et al. \cite{petto_interpreted_2024}, our proposed contributions eliminate the need to disclose process specifications to all involved participants, as seen in trust enhancing approaches for processes in related work \cite{weber_untrusted_2016,toldi_blockchain-based_2023}.
With our proposed setup, only the ZKP verification key for the last verifiable activity needs to be revealed to verifying participants.
All other business process properties, the \textit{guest program}, and verifiable process must only be disclosed to the certification agency which ensures the process confidentiality. 
The process specification, metadata, and private data inputs are preserved from other participants.
This ensures the required \textit{integrity} and \textit{transparency} attributes for the confidential process. %

%% file: content/computation_algorithm.tex
\newcommand{\var}{\texttt}
\newcommand{\Parameter}[2]{\Statex $\triangleright$ \var{#1}: #2}
\begin{algorithm}[t]
\small
\caption{\small{Simplified Carbon Footprinting zkVM Guest Program}}\label{alg:example}
\begin{algorithmic}
\Parameter{previousImageId}{The \textit{ImageId} of the previous footprinting activity}
\Parameter{previousReceipt}{The \textit{Receipt} of the previous footprinting activity}
\Parameter{emissionFactor}{The emission factor of the utilized resource}
\Parameter{resourceAmount}{The total amount of the consumed resource for the activity}
\If{\textproc{verify\_zkp}(\var{previousVerificationKey}, \var{previousZKP})}
    \State $\var{totalPreviousEmissions} \gets \var{previousZKP.totalEmissions} $
    
\Else 

\Return \var{False} \Comment{Returns False if verifying the previous ZKP fails.}
\EndIf 

\State $\var{currentEmissions} \gets \textproc{compute\_emissions}(\var{emissionFactor}, \var{resourceAmount})$
    
\Return \var{currentEmissions} + \var{totalPreviousEmissions}
\end{algorithmic}\end{algorithm}

%% file: media/table.tex
\scriptsize
\begin{tabular}{c|cc|cc|cc}
\multirow{2}{*}{\begin{tabular}[c]{@{}c@{}}No. of\\ Activities\end{tabular}} &
  \multicolumn{2}{c|}{Single Step} &
  \multicolumn{2}{c|}{Composite} &
  \multicolumn{2}{c}{Chained} \\ \cline{2-7} 
   & sec  & MB   & sec    & MB            & sec    & MB            \\ \hline
1  & 6.25 & 0.24 & 6.71   & 0.24          & 6.63   & 0.24          \\
5  & 6.18 & 0.24 & 51.42  & 2.19          & 33.57  & 1.22          \\
8  & 6.66 & 0.24 & 93.37  & \textbf{3.65} & 54.22  & 1.95          \\
9  & 7.02 & 0.24 & 106.86 & \textbf{4.13} & 60.93  & 2.19          \\
10 & 6.4  & 0.24 & 120.51 & 4.62          & 67.61  & 2.43          \\
15 & 6.51 & 0.24 & 203.24 & 7.05          & 102.13 & 3.65          \\
16 & 6.49 & 0.24 & 220.12 & 7.54          & 109.02 & \textbf{3.89} \\
17 & 6.7  & 0.24 & 239.71 & 8.03          & 115.99 & \textbf{4.13} \\
20 & 6.48 & 0.24 & 304.87 & 9.48          & 138.27 & 4.86          \\
25 & 6.37 & 0.24 & 384.15 & 11.92         & 174.45 & 6.08          \\
30 & 6.87 & 0.24 & 464.04 & 14.35         & 211.52 & 7.3     
\vspace{2.5em}
\end{tabular}

%% file: content/conclusion.tex
\section{Conclusion and Future Work}\label{sec:conclusion}
This paper uses the example of verifiable carbon footprinting to present zkVMs as a service technology within business processes to enable trustworthy, verifiable computations while preserving private data and process confidentiality in a novel, blockchain-less approach.
We evaluate three approaches to verifiable processes on the example of carbon footprinting processes.
Our results demonstrate that our proposed architecture ensures the integrity and confidentiality of process executions at the expense of increased processing overheads.
Dedicated footprinting activities ensure the required transparency for PCFs, while chained ZKPs provide the necessary degree of data confidentiality and verifiability. 
Our prototypical implementation demonstrates the fundamental capability of proving and verification within WMEs. 
It reveals practical integration challenges, particularly for chained composite proofs due to large proof sizes.

For future work, we envision advancements that strengthen the approach's validity through verifiable data authenticity and a platform to facilitate a more holistic setup of ZKP technology among participants.
We envision extending our work to explore the implications of our proposal on inter-organizational process management practices.

Additionally, we anticipate integrating hierarchical process models and enabling conditional routing and dynamic process adaptations through verifiable path validations.
Complementary to this, the role of process control flow and process structure on the suitability of the presented modeling alternatives may be further explored.
Moreover, this work could be extended by utilizing zkVMs as a vessel to enable verifiable executions of choreographies.
Broader, we expect that ZKPs within BPM enable more verifiable process conformance for auditing and analysis purposes.

\subsubsection{Acknowledgements}
The authors have no competing interests to declare that are relevant to the content of this article.